\documentclass[pra,twocolumn,showpacs,superscriptaddress,a4paper]{revtex4}
\usepackage{bm,graphicx,amsmath}

\newcommand{\ket}[1]{ \left. | #1 \right\rangle }
\newcommand{\mean}[1]{\left\langle#1\right\rangle}

\begin{document}

\title{Quantum theory of a polarization phase-gate in an atomic tripod configuration}

\author{S. Rebi\'{c}}
\email[E-mail: ]{stojan.rebic@unicam.it}
\author{D. Vitali}
\author{C. Ottaviani}
\author{P. Tombesi}
\affiliation{INFM and Dipartimento di Fisica, Universit\`{a} di Camerino, I-62032 Camerino, Italy}
\author{M. Artoni}
\affiliation{INFM and Dipartimento di Chimica e Fisica dei Materiali, Universit\`{a} di Brescia, I-25133 Brescia, Italy}
\affiliation{European Laboratory for Non-linear Spectroscopy, via N. Carrara, I-50019 Sesto
Fiorentino, Italy}
\author{F. Cataliotti}
\affiliation{European Laboratory for Non-linear Spectroscopy, via N. Carrara, I-50019 Sesto Fiorentino, Italy}
\affiliation{INFM and Dipartimento di Fisica, Universit\`{a} di Catania, I-95124 Catania, Italy}
\author{R. Corbal\'{a}n.}
\affiliation{Departament de F\'{i}sica, Universitat Aut\`{o}noma de Barcelona, E-08193, Bellaterra, Spain}

\begin{abstract}
We present the quantum theory of a polarization phase-gate that can be realized in a sample of ultracold rubidium atoms driven into a tripod configuration. The main advantages of this scheme are in its relative simplicity and inherent symmetry. It is shown that the conditional phase shifts of order $\pi$ can be attained.
\end{abstract}


\maketitle

\section{Introduction \label{sec:intro}}

Giant noise-free optical nonlinearities have been extensively studied over the last decade. In particular, atomic coherence effects such as coherent population trapping (CPT)~\cite{Arimondo96} and electromagnetically induced transparency (EIT)~\cite{Boller91} have been identified as promising mechanisms for the production of large nonlinear susceptibilities, while significantly reducing absorption and spontaneous emission noise~\cite{Boller91}. These effects have been observed in a three-level $\Lambda$ atomic configuration where an incident probe, in the presence of a pump and under a strict two-photon resonance condition, does not essentially interact with the atomic sample. Optical nonlinearities can be produced if this resonance condition is somewhat disturbed, by
introducing either an additional energy level(s)~\cite{Schmidt96,Zubairy02}, or a mismatch between the probe and the pump field frequencies~\cite{Matsko03}. Improvements of many orders of magnitude with respect to conventional nonlinearities have indeed been observed~\cite{Wang01}, but are limited to the case of classical fields~\cite{Grangier98}.

In this paper we examine a four-level atomic \textit{tripod} configuration. This has often been used as an extension of a $\Lambda$ scheme, e.g., by Paspalakis who suggested its potential for nonlinear optical processes~\cite{Paspalakis02}. It should be mentioned that while Paspalakis considered the case of a single weak probe, we are adopting a setup with two weak fields namely a $probe$ and and a $trigger$ in the presence of a strong pump. The
pump creates EIT, and leads to a large cross phase modulation (XPM) between the probe and the trigger~\cite{Schmidt96}. Such a modulation can be used to achieve optical phase shifts of the order $\pi$ radians which can be attained through an alternative phase-gate scheme that is here illustrated. The binary information is encoded in the polarization degree of
freedom~\cite{Ottaviani03} of an incident probe and trigger pulse while the phase-gate mechanism relies on an enhanced cross-phase modulation between the two pulses. The gating mechanism will be examined for either semiclassical or quantum probe and trigger fields.The four atomic level \emph{tripod} configuration is relatively simple yet robust requiring good control over frequencies and intensities of the probe and trigger laser pulses. We should also recall that when compared with other schemes~\cite{Wang01,Resch02}, the present scheme can substantially reduce the experimental effort for its realizability.

The paper is organized as follows. In Sec.~\ref{sec:dress}, dressed states analysis is performed. Dressed states are then used to present a simple physical picture illustrating
the requirements for a strong XPM. In Sec.~\ref{sec:susceptibilities}, the expressions for linear and nonlinear susceptibilities are derived for a case of classical fields. In Sec.~\ref{sec:quantpt} we analyze essentially the same system, but with quantized probe and trigger fields, and find that in adiabatic limit, a symmetry of results in semiclassical and quantum adiabatic cases exist. Sec.~\ref{sec:qpg} discusses a phase gate operation. Finally, Sec.~\ref{sec:conclusion} summarizes the results of the preceding Sections.

\section{Dressed States of the Tripod System \label{sec:dress}}

\begin{figure}[t]
\includegraphics[scale=0.7]{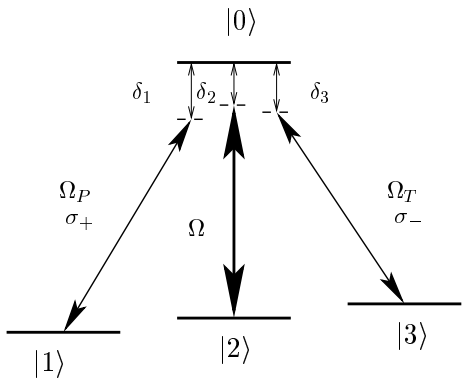}
 \caption{Energy level scheme for a tripod. Detunings $\delta_j= \omega_0-\omega_j-\omega_j^{(L)}$  denote the laser frequency ($\omega_j^{(L)}$) detunings from the respective transitions $|j\rangle \leftrightarrow |0\rangle$. States $|1\rangle$, $|2\rangle$ and $|3\rangle$ correspond to the states $|5S_{1/2}, F = 1, m = \{ -1, 0, 1 \} \rangle$ of $^{87}$Rb, while state $|0\rangle = |5P_{3/2}, F = 0 \rangle$. \label{fig:tripod} }
\end{figure}

The energy level scheme of a tripod system is given in Fig.~\ref{fig:tripod}. Probe and trigger fields have Rabi frequencies $\Omega_P$ and $\Omega_T$ and polarizations $\sigma_+$ and $\sigma_-$. The pump Rabi frequency is $\Omega$.
Among the four eigenstates in the system.~\cite{Unanyan98} two contain no contribution from the excited state $|0\rangle$ and hence they are \textit{dark states}. For the special case of atomic detunings $\delta_j = \delta = 0$, these are
\begin{subequations}
\label{eq:darkstates}
\begin{eqnarray}
    |e_1\rangle &=& \frac{\Omega_T |1\rangle -
\Omega_P|3\rangle}{\sqrt{\Omega_P^2+\Omega_T^2}} \label{eq:e1dark}, \\
    |e_2\rangle &=& \frac{\Omega \Omega_P |1\rangle + \Omega \Omega_T
|3\rangle - \left( \Omega_P^2 + \Omega_T^2 \right)
|2\rangle}{\sqrt{\left( \Omega_P^2+\Omega_T^2\right)\left(
\Omega_P^2+\Omega^2+\Omega_T^2 \right)}}.
\end{eqnarray}
\end{subequations}
The two other states,  
\begin{equation}
    |e_\pm\rangle = \frac{\Omega_P |1\rangle \pm |0\rangle + \Omega_T
|3\rangle + \Omega |2\rangle}{\sqrt{\Omega_P^2+\Omega^2+\Omega_T^2}}
\label{eq:brightstates}
\end{equation}
contain $\ket{0}$ and have energies $\pm\sqrt{\Omega_P^2+\Omega^2+\Omega_T^2}$. Necessary conditions for achieving a large cross-Kerr phase shift can be formulated as follows: $(i)$ probe and trigger must be tuned to dark states, $(ii)$ the transparency frequency window for each of these dark states has to be narrow and with a steep dispersion to enable significant group velocity reduction, and $(iii)$ there must be a degree of symmetry between the two transparency windows so that trigger and probe group velocities can be made to be equal.~\cite{Lukin00,Ottaviani03} These conditions can be satisfied by taking all three detunings nearly equal. When detunings differ by very small amounts yet falling within the
width of the transparency window, then strong cross-Kerr modulation with group velocity matching can still be achieved and phase gate operation realized.

\section{Semiclassical Susceptibilities \label{sec:susceptibilities}}

The expressions for the probe and trigger susceptibilities are obtained by solving the relevant optical Bloch equations~\cite{Rebic03} for the elements of density matrix $\rho$.
When $|\Omega|^2 \gg |\Omega_{P,T}|^2$ and $\Omega_P \approx \Omega_T$, the steady-state
population distribution will be symmetric, i.e., $\rho_{11} \approx \rho_{33} \approx \frac{1}{2}$, with vanishing population in the other two levels. The resulting general expression for the steady--state ($ss$) probe susceptibility is,
\begin{subequations}
\begin{eqnarray}
\chi_P &=& -4\pi\mathcal{N}|\bm{\mu}_P|^2 \, \frac{\left(\rho_{10}\right)_{ss}}{\Omega_P} ,\\
\chi_T &=& -4\pi\mathcal{N}|\bm{\mu}_T|^2 \, \frac{\left(\rho_{30}\right)_{ss}}{\Omega_T} .
\end{eqnarray}
\end{subequations}
Here $\mathcal{N}$ denotes the atomic density, $\Omega_{P,T} = -\left(\bm{\mu}_{P,T} \cdot \bm{\varepsilon}_{P,T}\right) E_{P,T}/\hbar$ the probe and trigger Rabi frequencies, with
$\bm{\varepsilon}_{P,T}$ and $\bm{\mu}_{P,T}$ being respectively their polarization unit vector and electric dipole matrix elements. The complex detunings
\begin{subequations}
\label{eq:gendet}
\begin{eqnarray}
\Delta_{j0} &=& \delta_j +i\gamma_{j0} ,\\
\Delta_{kj} &=& \delta_j - \delta_k - i\gamma_{kj} \ \ \  (k,j = 1,2,3).
\end{eqnarray}
\end{subequations}
are defined in terms of the general rate $\gamma_{kj}$ whereby $(k=j)$ and $(k\neq j)$ describe respectively population decays and collisional dephasings. In particular, $\gamma_{j0}$ denotes the decay of atomic coherences. We here assume equal decay and
dephasing rates $\gamma_{j0} = \gamma$, $\gamma_{kj} = \gamma_d$.

To examine cross-phase modulation between the probe and the trigger fields, it is sufficient to keep only the two lowest order contributions in the fields, so that
\begin{subequations}
\begin{eqnarray}
\chi_{P} &=& \chi_{P}^{(1)} + \chi_{P}^{(3)} |E_{T}|^2, \\
\chi_{T} &=& \chi_{T}^{(1)} + \chi_{T}^{(3)} |E_{P}|^2.
\end{eqnarray}
\end{subequations}
The linear susceptibilities
\begin{subequations}
\begin{eqnarray}
  \chi_P^{(1)} &=& \frac{\mathcal{N}|\bm{\mu}_P|^2}{\hbar\epsilon_0} \times \frac{1}{2}\frac{\Delta_{12}}{\Delta_{10}\Delta_{12}-|\Omega|^2}, \label{eq:linsuscp} \\
  \chi_T^{(1)} &=& \frac{\mathcal{N}|\bm{\mu}_T|^2}{\hbar\epsilon_0} \times \frac{1}{2}\frac{\Delta_{23}^*}{\Delta_{30}\Delta_{23}^*-|\Omega|^2}, \label{eq:linsusct}
\end{eqnarray}
\end{subequations}
are completely symmetric with respect to exchange $1 \leftrightarrow 3$ and $P  \leftrightarrow T$ and so are the
third-order nonlinearities
\begin{subequations}
\label{eq:kerrsusc}
\begin{eqnarray}
  \chi_P^{(3)} &=& \frac{\mathcal{N}|\bm{\mu}_P|^2|\bm{\mu}_T|^2}{\hbar^3\epsilon_0} \times \frac{1}{2}\frac{\Delta_{12}/\Delta_{13}}{\Delta_{10}\Delta_{12}-|\Omega|^2}\nonumber \\
&\ & \times \left( \frac{\Delta_{12}}{\Delta_{10}\Delta_{12}-|\Omega|^2}+\frac{\Delta_{23}}{\Delta_{30}^*\Delta_{23}-|\Omega|^2} \right), \label{eq:kerrsuscp} \\
  \chi_T^{(3)} &=& \frac{\mathcal{N}|\bm{\mu}_T|^2|\bm{\mu}_P|^2}{\hbar^3\epsilon_0} \times \frac{1}{2}\frac{\Delta_{23}^*/\Delta_{13}^*}{\Delta_{30}\Delta_{23}^*-|\Omega|^2} \nonumber \\
&\ & \times \left(
\frac{\Delta_{12}}{\Delta_{10}^*\Delta_{12}^*-|\Omega|^2}+\frac{\Delta_{23}^*}{\Delta_{30}\Delta_{23}^*-|\Omega|^2}
\right) . \label{eq:kerrsusct}
\end{eqnarray}
\end{subequations}
with $\chi_P^{(3)} \approx \chi_T^{(3)} = \chi^{(3)}$ for approximately equal atomic detunings. When all detunings $\delta_j$ acquire the same value, probe and trigger share a
strong control field with Rabi frequency $\Omega$ and exhibit all effects associated with EIT including strong group velocity reductions. It was first pointed out by Lukin and
Imamo\u{g}lu~\cite{Lukin00} appreciable nonlinear interaction can be achieved when the optical pulses interact within the transparent nonlinear medium for a sufficiently long time. This happens when both pulses group velocities are very small and equal, a requirement which can easily be met in our proposal owing to the natural symmetry of a tripod configuration.  Details can however be found in Rebi\'{c} {\em et al.}~\cite{Rebic03}.

\section{Quantized probe and trigger fields \label{sec:quantpt}}

We consider in this section the case in which probe and trigger are quantum fields. The pump is still considered much stronger than both of them so as to neglect its quantum fluctuations as for a classical field. We specifically adopt a recently developed formalism~\cite{Lukin00,Fleischhauer02} and apply it to our tripod atomic configuration. The relevant interaction Hamiltonian is
\begin{eqnarray}
H_{int} &=& -\int \frac{dz}{L} N \left[ \hbar \delta_1 \sigma_{00} + \hbar (\delta_1-\delta_2) \sigma_{22} + \hbar (\delta_1-\delta_3) \sigma_{33} \right. \nonumber \\
&\ &+ \hbar g_P \left(\hat{E}_P\sigma_{01} + \hat{E}_P^\dagger \sigma_{10}\right) + \hbar g_T \left(\hat{E}_T\sigma_{03} + \hat{E}_T^\dagger \sigma_{30}\right) \nonumber \\
&\ &\left . + \hbar \Omega (\sigma_{02} + \sigma_{20}) \right] , \label{eq:qham}
\end{eqnarray}
where $L$ is the interaction length along the propagation axis $z$, $g_{P,T}$ denotes coupling strengths of probe and trigger fields $\hat{E}_{P,T}$ to the respective atomic transitions, and $\sigma_{ij} = \frac{1}{N}\sum_k \sigma_{ij}^{(k)}$ are the collective operators for the atomic populations and transitions for $N$ atoms in a medium. The equations for probe and trigger pulses propagating along $z-$axis through the tripod-media are given by
\begin{subequations}
\begin{eqnarray}
\left( \frac{\partial}{\partial t} + c \frac{\partial}{\partial z} \right)\hat{E}_P (z,t) &=& -i g_P N \sigma_{10}, \label{eq:fieldeqP} \\
\left( \frac{\partial}{\partial t} + c \frac{\partial}{\partial z} \right)\hat{E}_T (z,t) &=& -i g_T N \sigma_{30}, \label{eq:fieldeqT}
\end{eqnarray}
while the equations for the atomic transition operators
\begin{eqnarray}
\dot{\sigma}_{10} &=& -i\Delta_{10}\sigma_{10} - i g_P \hat{E}_P (\sigma_{11}-\sigma_{00}) \nonumber \\
&\ &- i g_T \hat{E}_T \sigma_{13} - i \Omega \sigma_{12}, \\
\dot{\sigma}_{20} &=& -i\Delta_{20}\sigma_{20} - i g_P \hat{E}_P \sigma_{21} - i g_T \hat{E}_T \sigma_{23} \nonumber \\
&\ &- i \Omega (\sigma_{22} - \sigma_{00}) , \\
\dot{\sigma}_{30} &=& -i\Delta_{30}\sigma_{30} - i g_P \hat{E}_P \sigma_{31} \nonumber \\
&\ &- i g_T \hat{E}_T (\sigma_{33} - \sigma_{00}) - i \Omega \sigma_{32}, \\
\dot{\sigma}_{12} &=& -i\Delta_{12}\sigma_{12} + i g_P \hat{E}_P \sigma_{02} - i \Omega \sigma_{10}, \\
\dot{\sigma}_{13} &=& -i\Delta_{13}\sigma_{13} + i g_P \hat{E}_P \sigma_{03} - i \hat{E}_T^\dagger \sigma_{10}, \\
\dot{\sigma}_{23} &=& -i\Delta_{23}\sigma_{23} - i g_T \hat{E}_T^\dagger \sigma_{20} + i \Omega \sigma_{03}.
\end{eqnarray}
\end{subequations}
with generalized detunings defined as in Eqs.~(\ref{eq:gendet}).

We now proceed by assuming the low intensity probe and trigger, $g_j \langle \hat{E}_j \rangle \ll \Omega$ and strong pump $|\Omega|^2/\gamma_{0j}\gamma_{ij} \gg 1$. The latter
condition also implies that the EIT resonances for both, probe and trigger fields are strongly saturated. Furthermore, if $g_P \sim g_T$, for a probe and trigger fields of equal mean amplitudes $\langle \hat{E}_j \rangle$, we can assume $\langle \sigma_{00} \rangle \approx \langle \sigma_{22} \rangle \approx 0$ and $\langle \sigma_{11} \rangle \approx \langle \sigma_{33} \rangle \approx \frac{1}{2}$, as in the semiclassical case illustrated in the previous section. Under these conditions one arrives at the following equations for the pulse propagation
\begin{subequations}
\begin{eqnarray}
\left( \frac{\partial}{\partial t} + c \frac{\partial}{\partial z} \right)\hat{E}_P (z,t) &\cong& \frac{g_P N}{\Omega} \left( \frac{\partial}{\partial t} + i\Delta_{12} \right)\sigma_{12} , \\
\left( \frac{\partial}{\partial t} + c \frac{\partial}{\partial z} \right)\hat{E}_T (z,t) &\cong& \frac{g_T N}{\Omega} \left( \frac{\partial}{\partial t} - i\Delta_{23} \right)\sigma_{32} .
\end{eqnarray}
\end{subequations}
Note that these last two equations emphasize the full symmetry between the probe and the trigger dynamics, a symmetry which is intimately linked with the atomic population being equally distributed between levels $\ket{1}$ and $\ket{3}$.

We assume that the derivatives of the atomic operators vary slowly compared with their respective decay rates. Furthermore, if the Rabi frequencies of the probe and trigger quantum fields are much smaller than $\Omega$, we can perform the adiabatic elimination of
the atomic variables, to obtain
\begin{subequations}
\label{eq:propeqs}
\begin{eqnarray}
\left( \frac{\partial}{\partial z} + \frac{1}{v_g^{(P)}}
\frac{\partial}{\partial t} \right)\hat{E}_P &=& -\kappa_P \hat{E}_P +
 \beta_P \frac{\partial^2}{\partial t^2} \hat{E}_P \nonumber \\
&\ &+ i\eta_P \hat{E}_T^\dagger \hat{E}_T\hat{E}_P , \\
\left( \frac{\partial}{\partial z} + \frac{1}{v_g^{(T)}}
\frac{\partial}{\partial t} \right)\hat{E}_T &=& -\kappa_T
\hat{E}_T
+ \beta_T \frac{\partial^2}{\partial t^2} \hat{E}_T\nonumber \\
&\ &+ i\eta_T \hat{E}_T \hat{E}_P^\dagger \hat{E}_P ,
\end{eqnarray}
\end{subequations}
where group velocities of probe and trigger pulses are given in terms of the respective refraction indices $n_g^{(P,T)}$~\cite{Ottaviani03}
\begin{subequations}
\begin{eqnarray}
v_g^{(P,T)} &=& \frac{c}{1+n_g^{(P,T)}} , \\
n_g^{(P)} &=& \frac{1}{2} \frac{g_P^2 N}{\Delta_{10}\Delta_{12}-|\Omega|^2} , \\
n_g^{(T)} &=& \frac{1}{2} \frac{g_T^2 N}{\Delta_{30}\Delta_{23}^*-|\Omega|^2} .
\end{eqnarray}
Expressions for group velocities above are consistent with the semiclassical case~\cite{Rebic03} for the case of equal couplings $g_j$, equal detunings and weak probe and trigger. The low intensity condition $g_j \langle \hat{E}_j \rangle \ll \Omega$ has
its semiclassical counterpart in $|\Omega_{P,T}| \ll |\Omega|$. The two refractive indices $n_g^{(P,T)}$ further determine the transparency windows
\begin{equation}
 \Delta\omega_{tr}^{(P,T)} = \sqrt{\frac{c}{\gamma l}\frac{|\Omega|^2}{n_g^{(P,T)}}} ,
\end{equation}
where we assumed $\gamma_{j0} = \gamma$. In general case, $\gamma \rightarrow \gamma_{10}$ for probe and $\gamma \rightarrow \gamma_{30}$ for trigger. The wave dispersion coefficients
$\beta_{(P,T)}$ and the single-photon loss rates due to collisional dephasing are given respectively by
\begin{eqnarray}
\beta_P &=& \frac{\Delta_{10}^* n_g^{(P)}}{c|\Omega|^2} , \\
\beta_T &=& \frac{\Delta_{30} n_g^{(T)}}{c|\Omega|^2} .
\end{eqnarray}
\end{subequations}
and by $\kappa_i v_g^{(i)}$ where
\begin{subequations}
\begin{eqnarray}
\kappa_P &=& i\frac{\Delta_{12} n_g^{(P)}}{c} , \\
\kappa_T &=& i\frac{\Delta_{23}^* n_g^{(T)}}{c} , 
\end{eqnarray}
\end{subequations}
while the rates for nonlinear interaction between pulses in~(\ref{eq:propeqs}) are determined through the anharmonic coefficients
\begin{subequations}
\begin{eqnarray} 
\eta_P &=& \frac{l g_P^2 g_T^2 N}{2\pi c^2}
\times \frac{1}{2}
\frac{\Delta_{12}/\Delta_{13}}{\Delta_{10}\Delta_{12}-|\Omega|^2}
\nonumber \\
&\ &\times \left(
\frac{\Delta_{12}}{\Delta_{10}\Delta_{12}-|\Omega|^2}+\frac{\Delta_{23}}{\Delta_{30}^*\Delta_{23}-|\Omega|^2}
\right), \\
\eta_T &=& \frac{l g_P^2 g_T^2 N}{2\pi c^2} \times \frac{1}{2} \frac{\Delta_{23}^*/\Delta_{13}^*}{\Delta_{30}\Delta_{23}^*-|\Omega|^2}
\nonumber \\
&\ &\times \left(
\frac{\Delta_{12}^*}{\Delta_{10}^*\Delta_{12}^*-|\Omega|^2}+\frac{\Delta_{23}^*}{\Delta_{30}\Delta_{23}^*-|\Omega|^2}
\right) .
\end{eqnarray}
\end{subequations}

When single photon loss $\kappa_{P,T}$ and wave dispersion $\beta_{P,T}$ are negligible, the solution of these coupled equations can be written as
\begin{subequations}
\label{eq:fieldsols}
\begin{eqnarray}
\hat{E}_P (z,t) &=& \hat{E}_P (t') \exp{\left[ i\eta_P \hat{E}_T^\dagger (t') \hat{E}_T (t') \right]} , \\
\hat{E}_T (z,t) &=& \hat{E}_T (t') \exp{\left[ i\eta_T \hat{E}_P^\dagger (t') \hat{E}_P (t') \right]} ,
\end{eqnarray}
\end{subequations}
where $t' = t - z/v_g$. Following the approach of Lukin and Imamo\u{g}lu~\cite{Lukin00}, we conclude that for an initial \textit{multimode coherent states} $\ket{\alpha_P,\, \alpha_T}$ of frequency spread $\Delta\omega$, the probe and trigger fields after propagation through a sample of length $l$ become
\begin{subequations}
\begin{eqnarray}
&\, &\mean{\hat{E}_P (z,t)} = \alpha_P (t') \times \nonumber \\
&\ &\times \exp{\left\{\left[ -2\sin^2{(\Phi_P/2)} + i\sin{\Phi_P}\right]\frac{|\alpha_T (t')|^2}{\Delta\omega} \right\}} , \\
&\, &\mean{\hat{E}_T (z,t)} = \alpha_T (t') \times \nonumber \\
&\ &\times \exp{\left\{\left[ -2\sin^2{(\Phi_T/2)} + i\sin{\Phi_T}\right]\frac{|\alpha_P (t')|^2}{\Delta\omega} \right\}} .
\end{eqnarray}
\end{subequations}
where \begin{equation} \Phi_{P,T} = \left( \frac{c\eta_{P,T}}{l} \right) l \ \Delta\omega = c \ \eta_{P,T} \ \Delta\omega 
\end{equation} 
is the quantum phase shift obtained due to the nonlinear interaction between probe and trigger. It can be shown further that as a result of the nonlinear interaction the phase
shift $\Phi_{P,T}$ is acquired by a pair of probe and trigger single photon wave packets. This is a general property of a solutions for field operators given by Eqs.~(\ref{eq:fieldsols}), and has been proven by Lukin and Imamo\u{g}lu~\cite{Lukin00} (see also~\cite{Sanders92}). As shown by Petrosyan and Malakyan~\cite{Petrosyan04}, these large nonlinear phase shifts can be used to create quantum entanglement.

\section{Phase Gate Operation \label{sec:qpg}}

The phase gate operation has been studied in detail in~\cite{Rebic03}. This Section outlines the main points involved. We denote $\phi_0^{P,T} = k_{P,T}l$ vacuum phase shifts for probe and trigger respectively, $\phi_{lin}^{P,T} = k_{P,T}l (1+2\pi \chi_{P,T}^{(1)})$ linear phase shift and $\phi_{nlin}^{P,T}$ nonlinear phase shift in an atomic sample of length $l$. Then, for polarizations $\sigma^+$ and $\sigma^-$ for probe and trigger
\begin{subequations}
\begin{eqnarray}
  |\sigma^-\rangle_P |\sigma^-\rangle_T &\rightarrow& e^{-i(\phi_0^P+\phi_{lin}^T)}|\sigma^-\rangle_P |\sigma^-\rangle_T, \\
  |\sigma^-\rangle_P |\sigma^+\rangle_T &\rightarrow& e^{-i(\phi_0^P+\phi_0^T)}|\sigma^-\rangle_P |\sigma^+\rangle_T, \\
  |\sigma^+\rangle_P |\sigma^+\rangle_T &\rightarrow& e^{-i(\phi_{lin}^P+\phi_0^T)}|\sigma^+\rangle_P |\sigma^+\rangle_T, \\
  |\sigma^+\rangle_P |\sigma^-\rangle_T &\rightarrow& e^{-i(\phi_+^P+\phi_-^T)}|\sigma^+\rangle_P |\sigma^-\rangle_T,
\end{eqnarray}
\end{subequations}
and a conditional phase shift being $\phi = \phi_+^P + \phi_-^T - \phi_{lin}^P - \phi_{lin}^T$, with $\phi_+^P = \phi_{lin}^P + \phi_{nlin}^P$ and $\phi_-^T = \phi_{lin}^T + \phi_{nlin}^T$. Notice that only the nonlinear part contributes to the conditional
phase shift. Also notice that levels $|1\rangle$ and $|3\rangle$ are not energy degenerate therefore the role of probe and trigger cannot be reversed when their polarizations are exchanged. For a Gaussian trigger (probe) pulse of time duration $\tau_{T (P)}$, whose peak Rabi frequency is $\Omega_{T (P)}$, moving with group velocity $v_g^{T (P)}$ through the atomic sample, the nonlinear probe (trigger) phase shift is~\cite{Ottaviani03}
\begin{subequations}
\begin{eqnarray}
  \phi_{nlin}^{P} &=& k_{P}l \frac{\pi^{3/2}\hbar^2|\Omega_{T}|^2}{4|\bm{\mu}_{T}|^2}\, \frac{\rm{erf}[\zeta_{P}]}{\zeta_{P}} \, {\rm Re}[\chi_{P}^{(3)}],\\
 \phi_{nlin}^{T} &=& k_{T}l \frac{\pi^{3/2}\hbar^2|\Omega_{P}|^2}{4|\bm{\mu}_{P}|^2}\, \frac{\rm{erf}[\zeta_{T}]}{\zeta_{T}} \, {\rm Re}[\chi_{T}^{(3)}],
\end{eqnarray}
\end{subequations}
and
\begin{subequations}
\begin{eqnarray}
\zeta_{P} &=& (1-v_g^{P}/v_g^{T})\sqrt{2}l/v_g^{P}\tau_{T} , \\
\zeta_{T} &=& (1-v_g^{T}/v_g^{P})\sqrt{2}l/v_g^{T}\tau_{P} .
\end{eqnarray}
\end{subequations}
For $\Omega_P \approx \Omega_T = \gamma$, $\Omega = 4.5 \gamma$, and detunings $\delta_1 = 10.01\gamma$, $\delta_2 = 10\gamma$, $\delta_3 = 10.02\gamma$, by assuming a low dephasing rate $\gamma_d = 10^{-2} \gamma$, we obtain a conditional phase shift of $\pi$ radians. This result holds for a cold rubidium gas~\cite{Ottaviani03,Rebic03} of length $l = 0.7$ cm at a density ${\mathcal N} = 3 \times 10^{12}$ cm$^{-3}$ and a probe resonant wavelength $\lambda = 795$ nm. Electric dipole moments for both probe and trigger transitions are
 $|\bm{\mu}_{P,T}| \sim 10^{-29}$ Cm. With these parameters, group velocities are essentially the same, so that $\rm{erf}[\zeta_P] / \zeta_P \simeq \rm{erf}[\zeta_T] / \zeta_T \approx 2 / \sqrt{\pi}$.

\begin{figure}[t]
 \includegraphics[width=0.45\textwidth,height=0.4\textwidth]{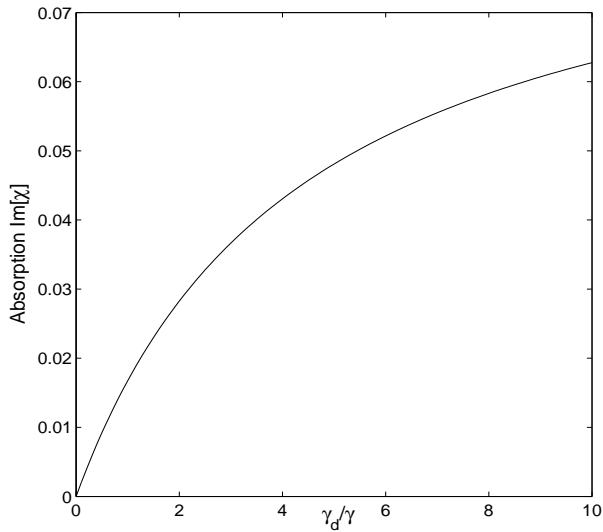}
 \caption{Probe absorption (scaled) at the center of probe transparency window, plotted against the dephasing rate, for $\Omega_P = \Omega_T = \gamma$, $\Omega = 4.5\gamma$, $\delta_j = 0$. \label{fig:abs}}
\end{figure}

Finally, we briefly comment on the role of fluctuations and dephasing. The laser fields intensity fluctuations generally increase fluctuations of the phase shift. In particular, intensity fluctuations of 1\% yield an error probability of about 4\% in the gate fidelity. This error enters through the fluctuations of the phase shifts $\phi_{P,T}^{lin}$ and $\phi_{P,T}^{nlin}$. We also assume that all lasers are tightly phase locked to each other in order to minimize the effects of relative detuning fluctuations. Dephasing, on the other hand, removes the singularity of nonlinear susceptibilities, but it also increases absorption as shown in Fig.~\ref{fig:abs}. For realistic parameter values of $\gamma_d \sim 10^{-2}\gamma$, this increase is seen to be negligible. It should be added that this conclusion holds for control fields of order $\Omega \sim \gamma$ and greater. For weaker control fields, the dephasing must also be lower.

\section{Conclusion \label{sec:conclusion}}
We have studied the nonlinear cross-kerr response between a probe and trigger field in a
four-level atomic sample driven in a tripod configuration. In particular, this has been studied for the case in which both probe and trigger are quantum fields evaluating their relevant phase sifts $\Phi_{P,T}$ when both probe and trigger are prepared in a multimode \textit{coherent} state. The theory can further be extended to evaluate the shift acquired by a pair of probe and trigger \textit{single-photon} wave packets. The large cross-Kerr
modulation between probe and trigger that can be obtained in the tripod configuration enables one to implement a phase gate with a conditional phase shift of the order of $\pi$. These large nonlinear phase shifts can also be used to create quantum entanglement~\cite{Petrosyan04}. The experimental feasibility of our tripod scheme has been assessed through a detailed study of the requirements needed to observe such a large shift in a realistic sample of magnetically trapped ultracold $\ ^{87}$Rb atoms.

\begin{acknowledgements}
We acknowledge enlightening discussions with P. Grangier, M.
Inguscio and M. Giuntini. We greatly acknowledge support from the
MURST (\emph{Actione Integrada Italia-Spagna}), the MIUR (PRIN
2001 \emph{Quantum Communications with Slow Light}) and by MCyT
and FEDER (\emph{project BFM2002-04369-C04-02}).
\end{acknowledgements}

\end{document}